\documentclass[pra,showpacs,showkeys,twocolumn,amsmath,amssymb,superscriptaddress,preprintnumbers,floatfix]{revtex4-1}
\usepackage{bm,amsfonts}
\usepackage{graphicx,color, wasysym}

\begin{document}
\renewcommand{\arraystretch}{2.3}

\title{Exact and explicit probability densities for one-sided L\'{e}vy stable distributions}

\author{K. A. Penson}
\email{penson@lptl.jussieu.fr}

\affiliation{Laboratoire de Physique Th\'{e}orique de la Mati\`{e}re Condens\'{e}e (LPTMC),
Universit\'{e} Pierre et Marie Curie, CNRS UMR 7600,
Tour 13 - 5i\`{e}me \'et., B.C. 121, 4 pl. Jussieu, F 75252 Paris Cedex 05, France}

\author{K. G\'{o}rska$^{1, \,}$}
\email{kasia\_gorska@o2.pl}

\affiliation{Nicolaus Copernicus University, Institute of Physics, ul. Grudzi\c{a}dzka 5/7,
PL 87-100 Toru\'{n}, Poland}


\begin{abstract}
We study functions $g_{\alpha}(x)$ which are one-sided, heavy-tailed L\'{e}vy stable probability distributions of index $\alpha$, $0<\alpha<1$, of fundamental importance in random systems, for anomalous diffusion and fractional kinetics. We furnish exact and explicit expressions for $g_{\alpha}(x)$, $0 \leq x < \infty$, for all $\alpha \,=\, l/k < 1$, with $k$ and $l$ positive integers. We reproduce all the known results given by $k\leq 4$ and present many new exact solutions for $k > 4$, all expressed in terms of known functions. This will allow a 'fine-tuning' of $\alpha$ in order to adapt $g_{\alpha}(x)$ to a given experimental situation.       
\end{abstract}

\maketitle

Theoretical description of many collective physical systems which include special sort of disorder or randomness often requires a radical departure from classical diffusive behaviour. On the probabilistic level this signifies the appearance of distributions with non-conventional characteristics, like diverging mean and variance along with all integer moments different from zeroth one. In this context the prominent r\^{o}le plays the discovery of particular distributions with these properties, called now L\'{e}vy stable laws \cite{JPKahane95}, whose generic example is $g_{1/2}(x)~=~(2\sqrt{\pi}\, x^{3/2})^{-1}\, \exp(-1/4x)$, $x\geq 0$; the word "stable" means here that the product of characteristic functions (cf) of two such laws is a cf of another law of the same type \cite{JPKahane95}. The general distribution of that type $g_{\alpha}(x)$ can be shown to possess the cf or the Laplace transform of the form \cite{JMikusinski59, HPollard46, WRSchneider86}:
\begin{equation}\label{eq1}
\int_{0}^{\infty} e^{-p\,x}\, g_{\alpha}(x)\, dx \,=\, e^{-p^{\alpha}}, \quad p>0\, , \, 0<\alpha<1,
\end{equation}
which is the well-known Kohlrausch-Williams-Watts function \cite{RSAnderssen04} or stretched exponential. Several independent proofs can be given that $g_{\alpha}(x)$ obeying Eq.~(\ref{eq1}) is positive \cite{JMikusinski59, HPollard46, IMSokolov00}.

The functions $g_{\alpha}(x)$ are ubiquitous in many fields of condensed and soft matter physics \cite{RMetzler04, PGDeGennes02, RHilfer02}, geophysics \cite{OSottolongo00}, meteorology \cite{MLagha07}, economics \cite{RNMantegna95}, fractional kinetics \cite{IMSokolov02, AVChechkin08}, etc. For instance, the value $\alpha = 1/4$ is thought to describe mechanical and dielectric properties of glassy polymers \cite{JTBendler84}. It is also confirmed that the same value of $\alpha$ is relevant for a statistical description of subrecoil laser cooling \cite{FBardon02, FBardou94}. In general, numerous phenomena falling in the class of subdiffusion \cite{TKoren07} call for $g_{\alpha}(x)$, $\alpha < 1$, in their theoretical description. On theoretical side, the L\'{e}vy stable distributions are essential tools in the study of random maps and resulting combinatorial structures \cite{PFlajolet}. The actual use of L\'{e}vy stable type distributions has been hampered for subjective and objective reasons \cite{VVUchaikin99, VVUchaikin03}. The subjective ones include a certain reticence to use distributions with both mean and variance diverging. The main objective reason is a lack of knowledge of $g_{\alpha}(x)$ for most values of $\alpha$. The existing interpolation formulae \cite{JTBendler84} appear to be cumbersome to use.

It seems that obtaining explicit $g_{\alpha}(x)$ for arbitrary $0~<~\alpha~<~1$ constitutes a true challenge: only for a limited number of values of $\alpha$, i.e. $\alpha = 1/2$ (s. $g_{1/2}(x)$ above), $1/4$ \cite{EBarkai01}, $1/3$ \cite{HScher75}, $2/3$ \cite{EWMontroll84}, and $3/4$ \cite{HScher75} the explicit forms of $g_{\alpha}(x)$ are known. The formal solution for arbitrary $\alpha$ \cite{WRSchneider86, RHilfer02} is only of a limited use as it requires series or asymptotic expansions, which may become problematic, especially for small $\alpha$. 

The objective of this work is to present an universal formula for $g_{\alpha}(x)$, $\alpha = l/k$, with $k>l$ positive integers, which is exact and explicit. It reproduces all the known cases enumerated above, and yields an infinity of new solutions for $k > 4$, of which we quote, for the first time, several instances.

Eq. (\ref{eq1}) for $\alpha = l/k$ can be inverted giving
\begin{equation}\label{eq2}
g_{l/k}(x) \,=\, \frac{\sqrt{k l}}{(2\pi)^{(k-l)/2}}\, \frac{1}{x}\, G^{k, 0}_{l, k}\left(\frac{l^{l}}{k^{k}\, x^{l}}\, \Big\vert\,^{\Delta(l, 0)}_{\Delta(k, 0)}\right),
\end{equation}
valid for all $x\geq 0$, where $G^{m, n}_{p, q}\left(z \vert^{(a_{p})}_{(b_{q})}\right)$ is the Meijer $G$ function \cite{OIMarichev83, APPrudnikov92} and $\Delta(k, a) = \frac{a}{k}, \frac{a+1}{k}, \ldots, \frac{a + k - 1}{k}$ is a special list of $k$ elements. Eq.~(\ref{eq2}) is listed without proof as a special case for $\nu = 0$ and $a = 1$ of formula 2.2.1.19, in vol. 5 of \cite{APPrudnikov92}. The detailed demonstration of Eq.~(\ref{eq2}) with a combined use of Laplace and Mellin transforms will be given elsewhere. It turns out that the r. h. s. of Eq.~(\ref{eq2}) is a finite sum of $k-1$ generalized hypergeometric functions of type $_{p}F_{q}\left(^{(a_{p})}_{(b_{q})}\big\vert z\right)$ \cite{APPrudnikov92}:
\begin{equation}\label{eq3}
g_{l/k}(x) = \sum_{j=1}^{k-1} \frac{b_{j}(k, l)}{x^{1+ j\,\frac{l}{k}}} \,\,_{l+1}F_{k}\left(^{1, \Delta(l,\, 1 + j\,l/k)}_{\quad\Delta(k, j+1)} \Big\vert (-1)^{k-l} \frac{l^l}{k^k\, x^l}\right),
\end{equation}
where $b_{j}(k, l)$ are numerical coefficients given by 
\begin{equation}\label{eq4}
b_{j}(k, l) = \frac{l^{j l/k}\,\sqrt{k l}}{k^j\,(2\pi)^{(k-l)/2}}\,  \frac{\left[\prod_{i=1}^{j-1}\Gamma\left(\frac{i-j}{k}\right)\right]\, \left[\prod_{i=j+1}^{k-1}\Gamma\left(\frac{i-j}{k}\right)\right]}{\prod_{i=1}^{l-1} \Gamma\left(\frac{i}{l}-\frac{j}{k}\right)},
\end{equation}
where $\Gamma(y)$ is Euler's gamma function. Eq.~(\ref{eq3}) is the exact implementation of the program outlined in the fundamental work of Scher and Montroll \cite{HScher75}, in which it was conjectured that $g_{l/k}(x)$ can be expressed in terms of $_{p}F_{q}$'s. However in \cite{HScher75} actually only one new instance $g_{3/4}(x)$ was written down. Our formula Eq.~(\ref{eq3}), after appropriate reductions in $_{p}F_{q}$'s, see below, gives all exactly known cases mentioned above \cite{VVUchaikin99, VVUchaikin03, EBarkai01, HScher75}, with $g_{1/4}(x)$ \cite{EBarkai01}
\begin{eqnarray}\label{eq5}
g_{1/4}(x) &=& \frac{b_{1}(4, 1)}{x^{5/4}}\,_{0}F_{2}\left(^{\,\,\,-}_{1/2, 3/4}\Big\vert \frac{-1}{4^4 x}\right) + \frac{b_{2}(4, 1)}{x^{3/2}}\\[0.7\baselineskip]\nonumber
&\times& \,_{0}F_{2}\left(^{\,\,\,-}_{3/4, 5/4}\Big\vert \frac{-1}{4^4 x}\right)
+ \frac{b_{3}(4, 1)}{x^{7/4}} \,_{0}F_{2}\left(^{\,\,\,-}_{5/4, 3/2}\Big\vert\frac{-1}{4^4 x}\right)
\end{eqnarray}
and offers an unlimited number of new solutions for $g_{l/k}(x)$, $k > 4$, e.g.:
\begin{eqnarray}\label{eq6}
g_{p/5}(x) = \sum_{j=1}^{4}\, \frac{b_{j}(5, p)}{x^{1 + j p/5}} \,_{p+1}F_{5}\left(^{1, \Delta(p, 1 + j\,p/5)}_{\Delta(5, j+1)} \Big\vert \frac{p^p}{5^5 x^p}\right), \quad
\end{eqnarray}
$p=1, \ldots, 4$, see Table \ref{tab1} for coefficients in Eqs.~(\ref{eq5}) and (\ref{eq6}), etc. 

\begin{table*}
\begin{tabular}{c| c c c c}
j & 1 & 2 & 3 & 4 \\ \hline
$b_{j}(4, 1)$ & $\frac{1}{4\, \Gamma\left(\frac{3}{4}\right)}$ & $\frac{-1}{4 \sqrt{\pi}}$ & $\frac{\sqrt{2}\,\Gamma\left(\frac{3}{4}\right)}{16\pi}$ & --- \\ 
$b_{j}(5, 1)$ & $\frac{\sqrt{5}\Gamma\left(\frac{1}{5}\right)}{20\pi\, B}$ & $\frac{-\sqrt{5}\Gamma\left(\frac{2}{5}\right)}{20\pi\, A}$ & $\frac{\sqrt{5}\Gamma\left(\frac{3}{5}\right)}{40\pi\, A}$ & $\frac{-\sqrt{5}\Gamma\left(\frac{4}{5}\right)}{120\pi\, B}$\\ 
$b_{j}(5, 2)$ & $\frac{\sqrt{5}\cdot 2^{2/5}\Gamma\left(\frac{1}{5}\right)}{10\sqrt{\pi}\Gamma\left(\frac{3}{10}\right)\, B}$ & $\frac{-\sqrt{5}\cdot 2^{4/5}\Gamma\left(\frac{2}{5}\right)}{10\sqrt{\pi}\Gamma\left(\frac{1}{10}\right)\, A}$ & $\frac{-\sqrt{5}\cdot 2^{1/5}\Gamma\left(\frac{3}{5}\right)}{100\sqrt{\pi}\Gamma\left(\frac{9}{10}\right)\, A}$ & $\frac{\sqrt{5}\cdot 2^{3/5}\Gamma\left(\frac{4}{5}\right)}{100\sqrt{\pi}\Gamma\left(\frac{7}{10}\right)\, B}$\\ 
$b_{j}(5, 3)$ & $\frac{3\sqrt{5}\cdot 3^{1/10} \Gamma\left(\frac{1}{5}\right)}{10 \Gamma\left(\frac{2}{15}\right) \Gamma\left(\frac{7}{15}\right)\, B}$ & $\frac{\sqrt{5}\cdot 3^{7/10} \Gamma\left(\frac{2}{5}\right)}{50 \Gamma\left(\frac{4}{15}\right) \Gamma\left(\frac{14}{15}\right)\, A}$ & $\frac{-3\sqrt{5}\cdot 3^{3/10} \Gamma\left(\frac{3}{5}\right)}{25 \Gamma\left(\frac{1}{15}\right) \Gamma\left(\frac{11}{15}\right)\, A}$ & $\frac{-7 \sqrt{5}\cdot 3^{9/10} \Gamma\left(\frac{4}{5}\right)}{750 \Gamma\left(\frac{8}{15}\right) \Gamma\left(\frac{13}{15}\right)\, B}$\\ 
$b_{j}(5, 4)$ & $\frac{4\cdot 2^{1/10}5^{-1/2}\sqrt{\pi}\Gamma\left(\frac{1}{5}\right)}{\Gamma\left(\frac{3}{10}\right) \Gamma\left(\frac{1}{20}\right) \Gamma\left(\frac{11}{20}\right)\, B}$ & $\frac{6\cdot 2^{7/10}5^{-3/2}\sqrt{\pi}\Gamma\left(\frac{2}{5}\right)}{\Gamma\left(\frac{1}{10}\right) \Gamma\left(\frac{7}{20}\right) \Gamma\left(\frac{17}{20}\right)\, A}$ & $\frac{14\cdot 2^{3/10}5^{-5/2}\sqrt{\pi}\Gamma\left(\frac{3}{5}\right)}{\Gamma\left(\frac{9}{10}\right) \Gamma\left(\frac{3}{20}\right) \Gamma\left(\frac{13}{20}\right)\, A}$ & $\frac{11\cdot 2^{9/10}5^{-7/2}\sqrt{\pi}\Gamma\left(\frac{4}{5}\right)}{\Gamma\left(\frac{7}{10}\right) \Gamma\left(\frac{9}{20}\right) \Gamma\left(\frac{19}{20}\right)\, B}$
\end{tabular}
\caption{\label{tab1} Coefficients of Eqs. (\ref{eq5}) and (\ref{eq6}); $A = \sin(\pi/5)$ and $B = \sin(2\pi/5)$.}
\end{table*}

The symbol $\Delta(k, a)$ in Eq.~(\ref{eq3}) permits one to encode all the possible cases of $k$ and $l$ in a single formula. However we draw attention to the fact that cancellations will appear there due to the obvious identity $_{p+r}F_{q+r}\left(^{(a_{p}), (\alpha_{r})}_{(b_{q}), (\alpha_{r})} \Big\vert\, x\right)= \,_{p}F_{q}\left(^{(a_{p})}_{(b_{q})} \Big\vert\, x\right)$, where $(\alpha_{r})$ is an arbitrary sequence of $r$ parameters not equal to zero or to negative integers. Thus Eq.~(\ref{eq5}) is a sum of three $_{0}F_{2}$ functions, and likewise $g_{3/4}(x)$, which is not specified here, will be a sum of three $_{2}F_{2}$ functions, neatly confirming Eq.~(C10) of \cite{HScher75}, etc. In this manner for any $l/k$, a closed form of $g_{l/k}(x)$ can be obtained from Eq.~(\ref{eq3}). However, only for $k\leq 3$ it can be written down in terms of standard special functions \cite{RSAnderssen04, HScher75, VVUchaikin99, VVUchaikin03, EWMontroll84}.

A heuristic indication of how Eq.~(\ref{eq3}) comes about can be obtained from the series representation for $g_{\alpha}(x)$ derived by Humbert \cite{PHumbert45}, discussed for example by Hughes \cite{BDHughes95} and used in \cite{PFlajolet}. The series
\begin{equation}\label{eq7}
g_{\alpha}(x) = \frac{1}{\pi}\, \sum_{j=1}^{\infty} \frac{(-1)^{j+1}}{j!\, x^{1+\alpha\,j}}  \, \Gamma(1+\alpha\, j)\, \sin(\pi\, \alpha\, j)
\end{equation}
is a convergent expansion valid for all $0<\alpha<1$ and $x>0$. For $\alpha = l/k$ the decomposition of the summation index $j$ modulo $k$ yields an equivalent representation of $g_{l/k}(x)$ as a sum of $k-1$ infinite series. The structure of the coefficients in each of these series involve gamma functions ratios of type $\Gamma(l\, i + \theta)/\Gamma(k\, i + \phi)$, where $i$ is the new summation index and $\theta$ and $\phi$ are simple functions of $k$ and $l$. The application of Gauss-Legendre multiplication formula to both of these gamma functions allows the identification of $\,_{p}F_{q}$'s in Eq.~(\ref{eq3}) and the extraction of coefficients $b_{j}(k, l)$ of Eq.~(\ref{eq4}).

The advantage of our solution, Eqs.~(\ref{eq3}) and (\ref{eq4}) over Eq.~(\ref{eq7}) is clearly seen in practice, in conjunction with the use of computer algebra systems \cite{maple1}. Since in recent versions of these systems the hypergeometric functions $\,_{p}F_{q}$ as well as the Meijer G function are fully implemented, their use permits high-precision calculations. For reader's convenience we give in \cite{syntax1} the Maple$^{\tiny{\textregistered}}$ syntax for $g_{l/k}(x)$, see Eq.~(\ref{eq2}) above. Our experience indicates that for small $\alpha$ our results for small $x$ are more practical to use than the $x\to 0$ asymptotics given in \cite{JMikusinski59}. The reason is that there the region of applicability of Mikusi\'{n}ski's asymptotic expansion \cite{JMikusinski59} shrinks to exceedingly small values of $x$. For example for $\alpha = 1/20$, $g_{1/20}(0) = 0$, but a huge peak in $g_{1/20}(x)$ appears already at $x\sim 10^{-14}$. In contrast, our formulae work fine for any $x$ in this region. In the opposite limit for $\alpha\apprle 1$ the Humbert expansion Eq.~(\ref{eq7}) is slowly convergent for $x<\alpha$ but approximation \cite{JMikusinski59} works well as then $g_{\alpha}(x)$ is very close to zero in a considerable region near $x=0$ (e.g. already for $\alpha = 5/6$ the function $g_{5/6}(x)$ is practically equal to zero up to $x \approx 0.35$). Such a practically flat region for small $x$ can also be seen for $\alpha = 4/5$, compare the curve III on Fig.~\ref{fig3}. 

\begin{figure}[h]
\begin{center}
\includegraphics[scale=0.4]{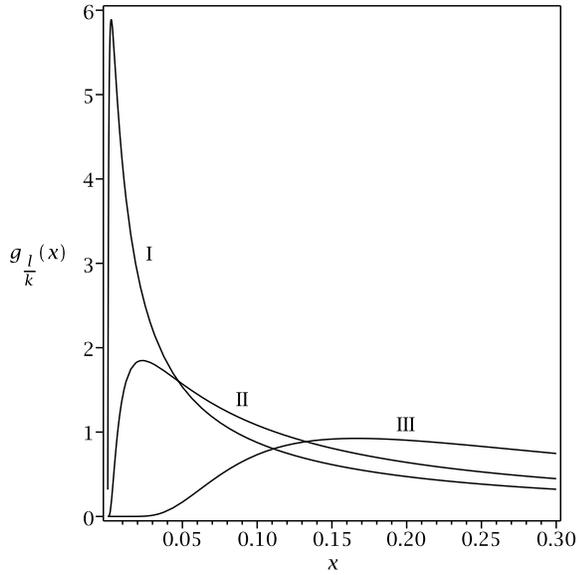}
\caption{\label{fig1} Comparison of $g_{l/k}(x)$: the curves I, II and III correspond to $l/k = 1/4$ (s. Eq.~(\ref{eq5})), $1/3$ and $1/2$, respectively.}
\end{center}
\end{figure}

\begin{figure}[h]
\begin{center}
\includegraphics[scale=0.4]{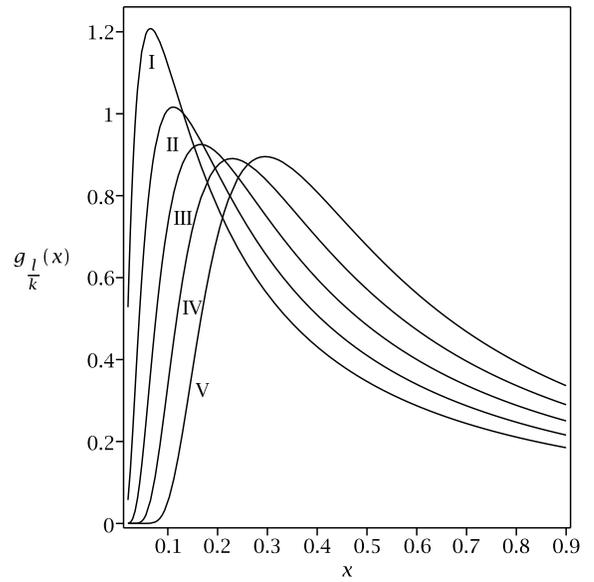}
\caption{\label{fig2} Comparison of $g_{l/k}(x)$: the curves I, II, III, IV and V correspond to $l/k = 2/5, 9/20, 1/2, 11/20$ and $3/5$, respectively. Calculations were performed using Eqs. (\ref{eq3}) and (\ref{eq4}).}
\end{center}
\end{figure}

\begin{figure}[h]
\begin{center}
\includegraphics[scale=0.4]{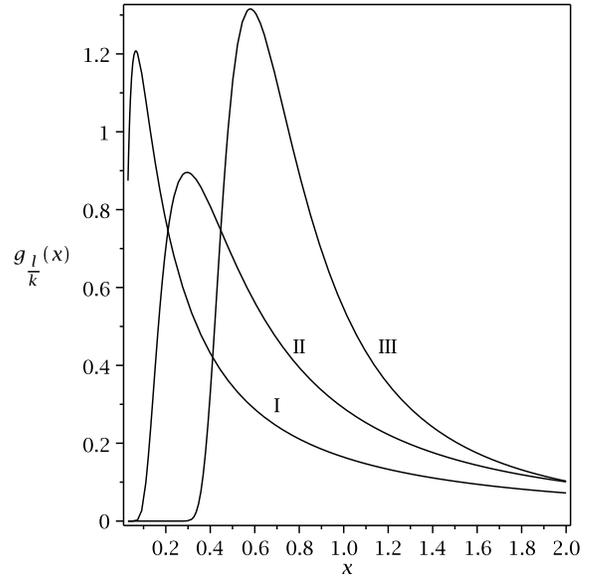}
\caption{\label{fig3} Comparison of $g_{l/k}(x)$: the curves I, II and III correspond to $l/k = 2/5, 3/5$ and $4/5$, respectively. Calculations were performed using Eq. (\ref{eq6}).}
\end{center}
\end{figure}

In Fig.~\ref{fig1} we compare three distributions for $l/k = 1/2$, $1/3$ and $1/4$. The salient feature for $l/k = 1/4$ is the appearance of a sharp maximum for very small $x$ so that these three curves can be barely shown on the same scale. Analogously, for $l/k = 1/5$ the maximum of $g_{1/5}(x)$ appears at $x_{0}(1/5) \approx 0.0002$ and the value $g_{1/5}(x_{0})\approx 25$. For $x<x_{0}(1/5)$ the values of $g_{1/5}(x)$ are very close to zero. As already mentioned above, for smaller values of $l/k$ this type of behaviour is even more pronounced and it explains \textit{a posteriori} the difficulties encountered in devising approximations valid for small $l/k$ and small $x$ \cite{EWMontroll84, JTBendler84}. In Fig.~\ref{fig2} we present the comparison of several distributions for values $l/k \approx 1/2$. Here the 'sharpening' of the distributions, as $l/k$ goes from $1/2$ to smaller values, is very clearly visible but is less dramatic than in Fig.~\ref{fig1}. We present in Fig.~\ref{fig3} the new distributions $g_{p/5}(x)$ given by Eq.~(\ref{eq6}) for $p = 2, 3$ and $4$.

All these probability distributions share the following features: (a) $g_{\alpha}(x)\to 0$, for $x~\to~0$, where they present an essential singularity $\sim x^\frac{-2+\alpha}{ 2(1-\alpha)} \exp[-A(\alpha) x^\frac{-\alpha}{1-\alpha}]$, $A(\alpha)>0$ \cite{JMikusinski59}; (b) $g_{\alpha}(x)\to B(\alpha) x^{-(1+\alpha)}$, for $x\to\infty$, $B(\alpha)>0$, indicating heavy-tailed asymptotics for large $x$; (c) all their fractional moments $M_{\alpha}(\mu)~=~\int_{0}^{\infty}x^{\mu} g_{\alpha}(x) dx = \Gamma(-\mu/\alpha)/[\alpha\, \Gamma(-\mu)]$, for real $\mu$, $-\infty<\mu<\alpha$, including $M_{\alpha}(0)=1$, are finite, and are infinite otherwise; (d) $g_{\alpha}(x)$ are unimodal with the maximum at $x_{0}(\alpha)$, and $x_{0}(\alpha)\to 0$ as $\alpha\to 0$.

The distributions $g_{\alpha}(x)$ constitute basic ingredients of all theories of anomalous diffusion where they are employed to produce solutions $P_{\alpha}(x, t)$ in space-time domain of various forms of the Fokker-Planck equations along with their fractional generalizations \cite{IMSokolov00, EBarkai01}. For instance in \cite{EBarkai01} $P_{\alpha}(x, t)$ is given as a convolution (called there inverse L\'{e}vy transform) of $\frac{d}{d s}\left[- g_{\alpha}(t/s^{1/\alpha})\right]$ with $P_{1}(x, t)$ being a normalized solution of the ordinary Fokker-Planck equation, see Eq.~(\ref{eq1}) in \cite{EBarkai01}. The explicit forms of $g_{\alpha}(x)$ presented here will permit further development of this ambitious approach.

The availability of $g_{l/k}(x)$ makes it possible to fully describe the long tail distributions of carrier transit times in amorphous materials such as As$_{2}$Se$_{3}$ and TNF-PVK. In fact in classic work \cite{HScher75} the measured values of $\alpha$ for these two materials were $\alpha = 0.45$ and $\alpha = 0.8$ respectively, compare Fig.~6 of \cite{HScher75}. These values were for a long time intractable theoretically. From now on, setting $\alpha = 9/20$ and $\alpha = 4/5$ in our Eqs.~(\ref{eq2}) and (\ref{eq3}) directly provides the sought for framework for interpretation of these data. The appropriate distributions are presented as the curve II in Fig.~(\ref{fig2}), ($\alpha = 9/20$) and the curve III in Fig.~(\ref{fig3}), ($\alpha = 4/5$).

We believe that the exact forms of $g_{\alpha}(x)$ obtained in this work, along with their asymptotics for $x\to\infty$ and exact values of fractional moments, constitute a solid basis to extract a value of $\alpha$ best suited for an experimental situation at hand. Once it has been done, such description can be further 'fine-tuned' by choosing values of $k$ and $l$ which would optimise the choice of $\alpha$. We hope that this approach will prove useful in practical applications.
\ \\

We thank Professor E. Barkai for kindly informing us about his results obtained in \cite{EBarkai01}. 

The authors acknowledge support from Agence Nationale de la Recherche (Paris, France) under Program No. ANR-08-BLAN-0243-2.

\ \\

\ \\

\begin{thebibliography}{10}

\bibitem{JPKahane95} J.-P.~Kahane, in \textit{L\'{e}vy Flights and Related Topics in Physics}, (Lecture Notes in Physics, vol. 450), edited by M.F. Shlesinger, G.M. Zaslavsky, and U. Frisch, (Springer, Berlin, 1995).

\bibitem{HPollard46} H.~Pollard, Bull. Amer. Math. Soc. \textbf{52}, 908 (1946).

\bibitem{JMikusinski59} J.~Mikusi\'{n}ski, Studia Math. \textbf{18}, 191 (1959).

\bibitem{WRSchneider86} W.~R.~Schneider, in \textit{Stochastic Processes in Classical and Quantum Systems} (Lecture Notes in Physics, vol. 262), edited by S.~Albeverio, G.~Casati, and D.~Merlini  (Springer, Berlin, 1986).

\bibitem{RSAnderssen04} R.~S.~Anderssen, S.~A.~Husain, and R.~J.~Loy, ANZIAM J. \textbf{45}, C800 (2004).

\bibitem{IMSokolov00} I.~M.~Sokolov, Phys. Rev. E \textbf{63}, 011104 (2000).

\bibitem{RMetzler04} R.~Metzler and J.~Klafter, J. Phys. A \textbf{37}, R161 (2004).

\bibitem{PGDeGennes02} P.-G.~de~Gennes, Macromol. \textbf{35}, 3785 (2002).

\bibitem{RHilfer02} R.~Hilfer, Phys. Rev. E \textbf{65}, 061510 (2002).

\bibitem{OSottolongo00} O.~Sottolongo-Costa, J.~C.~Antoranz, A.~Posadas, F.~Vidal, and A.~Vazquez, Geophys. Res. Lett. \textbf{27}, 1965 (2000).

\bibitem{MLagha07} M.~Lagha and M.~Bensebti, hal-00194153 (2007).

\bibitem{RNMantegna95} R.~N.~Mantegna and H.~E.~Stanley, Nature \textbf{376}, 46 (1995).

\bibitem{IMSokolov02} I.~M.~Sokolov, J.~Klafter, and A.~Blumen, Phys. Today \textbf{55}, 48 (2002).

\bibitem{AVChechkin08} A.~V.~Chechkin, V.~Yu.~Gonchar, R.~Gorenflo, N.~Korabel, and I.~M.~Sokolov, Phys. Rev. E \textbf{78}, 021111 (2008).

\bibitem{JTBendler84} J.~T.~Bendler, J. Stat. Phys. \textbf{36}, 625 (1984).

\bibitem{FBardou94} F.~Bardou, J.-P.~Bouchaud, O.~Emile, A.~Aspect, and C.~Cohen-Tannoudji, Phys. Rev. Lett. \textbf{72}, 203 (1994).

\bibitem{FBardon02} F.~Bardou, J.-P.~Bouchaud, A.~Aspect, and C.~Cohen-Tannoudji, \textit{L\'{e}vy Statistics and Laser Cooling} (Cambridge University Press, 2002).

\bibitem{TKoren07} T.~Koren, J.~Klafter, and M.~Magdziarz, Phys. Rev. E \textbf{76}, 031129 (2007).

\bibitem{PFlajolet} P.~Flajolet and R.~Sedgewick, \textit{Analytic Combinatorics} (Cambridge University Press, 2009).

\bibitem{VVUchaikin99} V.~V.~Ucha\u{i}kin, Zh. \'{E}ksp. Teor. Fiz. \textbf{115}, 2113 (1999) [Sov. Phys. JETP \textbf{88}, 1155 (1999)].

\bibitem{VVUchaikin03} V.~V.~Ucha\u{i}kin, Usp. Fiz. Nauk \textbf{173}, 847 (2003) [Sov. Phys. Usp. \textbf{46}, 821 (2003)].

\bibitem{EBarkai01} E.~Barkai, Phys. Rev. E \textbf{63}, 046118 (2001).  

\bibitem{HScher75} H.~Scher and E.~W.~Montroll, Phys. Rev. B \textbf{12}, 2455 (1975).

\bibitem{EWMontroll84} E.~W.~Montroll and J.~T.~Bendler, J. Stat. Phys. \textbf{34}, 129 (1984).

\bibitem{PHumbert45} P.~Humbert, Bull. Soc. Math. Fr. \textbf{69}, 121 (1945).

\bibitem{BDHughes95} B.~D.~Hughes, \textit{Random Walks and Random Environments}, vol. 1 (Clarendon Press, Oxford, 1995).

\bibitem{maple1} We have made extensive use of Maple$^{\tiny{\textregistered}}$ in this work.

\bibitem{syntax1} Here is the Maple$^{\tiny{\textregistered}}$ procedure {\ttfamily LevyDist(k, l, x)} used to calculate Eq.~(\ref{eq2}): \\
{\ttfamily LevyDist := proc(k, l, x) simplify(convert( sqrt(k * l) * MeijerG([[],[seq(j1/l, j1 = 0..l-1)]] , [[seq(j2/k, j2 = 0..k-1)],[]], l\textasciicircum l /(k\textasciicircum k * x\textasciicircum l)) /(x*(2*Pi)\textasciicircum ((k-l)/2)), StandardFunctions)); end;} .\\
Analogous syntax can be given for Mathematica$^{\tiny{\textregistered}}$.

\bibitem{OIMarichev83} O.~I.~Marichev, \textit{Handbook of Integral Transforms of Higher Transcendental Functions. Theory and Algorithmic Tables} (Ellis Horwood Ltd, Chichester, 1983).

\bibitem{APPrudnikov92} A.~P.~Prudnikov, Yu.~A.~Brychkov, and O.~I.~Marichev, \textit{Integrals and Series}, vols. 1-5 (Gordon and Breach, Amsterdam, 1992-1998).

\end{thebibliography}
\end{document}